# Mechanical Properties of Phagraphene Membranes: A Fully Atomistic Molecular Dynamics Investigation

J. M. de Sousa[1,2], A. L. Aguiar[2], E. C. Girão[2], Alexandre F. Fonseca[1], A. G. Sousa Filho[3], and Douglas S. Galvao[1,4]

[1]*Applied Physics Department, Institute of Physics "Gleb Wataghin", University of Campinas – UNICAMP, Campinas, São Paulo, CEP 13083-859, Brazil.*

[2]*Departamento de Física, Universidade Federal do Piauí, Teresina-PI, 64049-550, Brazil.*

[3]*Departamento de Física, Universidade Federal do Ceará, Fortaleza-CE, 60445-900, Brazil.*

[4]*Center for Computational Engineering and Sciences, UNICAMP, Campinas, São Paulo, Brazil.*

ABSTRACT

*Recently, a new 2D carbon allotrope structure, named phagraphene (PG), was proposed. PG has a densely array of penta-hexa-hepta-graphene carbon rings. PG was shown to present low and anisotropic thermal conductivity and it is believed that this anisotropy should be also reflected in its mechanical properties. Although PG mechanical properties have been investigated, a detailed and comprehensive study is still lacking. In the present work we have carried out fully atomistic reactive molecular dynamics simulations using the ReaxFF force field, to investigate the mechanical properties and fracture patterns of PG membranes. The Young's modulus values of the PG membranes were estimated from the stress-strain curves. Our results show that these curves present three distinct regimes: one regime where ripples dominate the structure and mechanical properties of the PG membranes; an elastic regime where the membranes exhibit fully planar configurations; and finally a plastic regime where permanent deformations happened to the PG membrane up to the mechanical failure or fracture.*

## INTRODUCTION

Recently, a new two-dimensional nanostructure composed of 5-6-7 rings of carbon atoms, named phagraphene (PG) [1], was proposed. This new form of carbon allotrope is predicted to have very interesting electromechanical properties, as compared to graphene [1]. PG presents anisotropic and relatively low thermal conductivity: in average of $218 \pm 20$ $Wm^{-1}K^{-1}$ along the armchair direction and $285 \pm 29$ $Wm^{-1}K^{-1}$ along the zigzag direction at room temperature [2]. Because of its distinct geometric arrangement of carbon atoms, PG exhibits interesting mechanical properties. A recently theoretical study performed using molecular dynamics (MD) methods with Tersoff empirical potential, showed that PG membranes have an elastic modulus of approximately 800 ± 14 GPa [2]. Another theoretical work showed that PG membranes might present other non-planar configurations [3], which can significantly influence PG mechanical properties. In spite of these studies a detailed and comprehensive study of PG mechanical properties is still missing and it is one of the objectives of the present work.

In this work, we carried out a detailed investigation (mechanical properites and fracture patterns) of PG membranes thorugh fully atomistic reactive molecular dyncamics (MD) simulations. For comparative purposes, we also considered similar graphene membranes.

**METHODOLOGY**

PG and graphene membranes were investigated under axial tension applied along two different directions: ($x$ and $y$ directions, see Figure 1a). We carried out MD simulations with the use of the reactive potential ReaxFF [4], as implemented in the LAMMPS code [5]. ReaxFF is a quantum mechanics based parameterized reactive force field. It can describe breaking and formation of covalent chemical bonds, thus being an ideal force field for the study of the fracture mechanics of the nanostructured systems [6-8].

PG and graphene membranes were initially thermalized at 300 K and this temperature was kept constant during the stretching dynamics until complete fracture. In order to estimate the thermal effects on the mechanical properties, we also carried out simulations at 900 K. The thermal equilibrium was performed through a canonical ensemble (NVT) with the use of the Nosé-Hoover thermostat [9], as implemented in the LAMMPS code [5]. The investigated membrane sizes were ~ $86 \times 86$ Å$^2$, corresponding to 2860 and 3024 carbon atoms for PG and graphene structures, respectively. We applied periodic boundary conditions along $x$ and $y$ directions. Along the $z$ direction, the MD simulation box size was fixed at a large size (~150 Å), in order to avoid spurious interactions between periodic images. In order to assure precise estimation of the mechanical properties under strain, before starting the axial stretching dynamics, we used a pressure barostat characterized by the isothermal-isobaric ensemble (NPT) [10], to release any internal stress at the edges and throughout the membrane making the total pressure equal to zero. The arrangement of the atoms is updated with each time step of molecular dynamics simulation of 0.05 fs. The dynamics of membrane stretching is performed by increasing the simulation box dimensions at a rate of $10^{-6}$/fs, which is sufficient to allow the system to equilibrate before the next MD step.

The modulus of elasticity is estimated through the stress-strain curve. The Young's modulus along $i$ direction is defined by $Y_i = \sigma_{ii}/\varepsilon_i$, where $\sigma_{ii}$ and $\varepsilon_i$ are the axial component of the stress tensor and strain along direction $i$, respectively [6-8]. The stress tensor is defined by $\sigma_{ij} = A^{-1}(\sum m_k v_{ki} v_{ki} + \sum m_k r_{ki} f_{ki})$, where the summations are on index $k$ from 1 to the total number of atoms $N$. $A$ is the area of membranes, $v_{ki}$ ($r_{ki}$) {$f_{ki}$} is the *i-esim* component of the velocity (position) {force per atom} vector of the *k-esim* carbon atom [6-8]. The stretching dynamics applied to the PG and graphene membranes until the complete fracture is configured in a molecular arrangement divided into two directions ($x$ and $y$). The descriptions of the fracture mechanics of the PG and graphene membranes are analysed through the von Mises stress tensor [6-8], by which we estimate the fracture pattern and the accumulated tension distribution within the membranes.

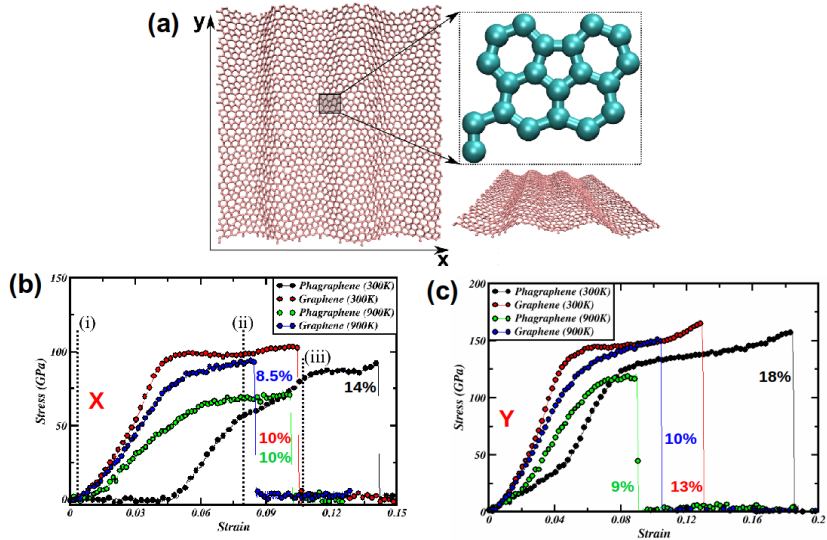

**Figure 1.** (a) Atomic model of a phagraphene (PG) membrane equilibrated in vacuum at room temperature of 300K. In the inset, a zoomed view of the PG unit cell with 20 atoms. (b) Stress-Strain curve for PG and graphene membranes with axial strain applied along the *x* direction at 300K and 900K, respectively. The curses typically exhibit three different regimes (i), (ii) and (iii) (see text for discussions). (c) Corresponding stress-strain curves for axial strain applied along the *y* direction. Critical strains are explicitly colored for each respective curve. Reader is referred to the color legend in the electronic version of the paper or identify the curves as follows: from top-left to bottom-right stress-strain curves: PG at 300 K, graphene at 300 K, PG at 900 K and graphene at 900 K.

**RESULTS**

In the following, we discuss the results of the structural dynamics, mechanical and fracture patterns for PG and graphene membranes. Firstly, we estimated the elastic modulus of PG and graphene membranes. Their Young's modulus, *Y*, was obtained from the analysis of the stress versus strain graphs, which can be visualized in Figures 1b and 1c. Our *Y* values were obtained within the linear region at the elastic limit (low strain) of the membranes, where for the graphene at 300K (900 K) we obtained 1131 (1220) and 1317 (1250) GPa along *x* and *y* directions, respectively. These values are in good agreement to the experimental results for the Yong's modulus of graphene reported in literature (1.0 ± 0.1 TPa [11]). In Figure 1c we present the corresponding values for strain applied along *y* direction. All curves exhibit basically two main mechanical regimes; elastic and plastic, which are characterized by the linear and planar regions of the stress-strain curves. Along *x* direction, graphene also presents the same two regime mechanical behavior before fracture (Figure 1b). However, for the PG membrane at 300 K, an interesting unusual elastic behavior occurs. As predicted by Podlivaev and Openov [3], ripples along the *x* direction were observed for PG, even in equilibrium state (see Figure 1a). These membrane ripples strongly influence the stress-strain curve along *x* direction, when compared to that along *y* direction (Figure 1c). As we can see in Figure

1b, for PG at 300 K (stress-strain black color or bottommost curve) a third distinct mechanical stress-strain regime appears (indicated by (i)). Within this region, we observe an unusual behavior, called here as a *pre-elastic* regime, where the structure can be easily strained from 0 to about 4.5% with very low stress, as compared to the other regimes or other strains along *y* direction (Figure 1c). This *pre-elastic* regime is not observed for PG strained along *y* direction and graphene, and it forms a very short region for PG at 900 K (green or second bottommost stress-strain curve in Figure 1b). Above ~ 4.5% of strain, the PG structure presents the regular elastic and plastic regimes ((ii) and (iii), respectively). This *pre-elastic* regime can be explained by the existence of the ripples, as we can stretch with low strain until the membranes undergo a transition to a quasi-planar configuration, thus entering the elastic regime.

The elastic regime is then observed as a linear behavior of the stress-strain curves. The dotted vertical line ((ii) in Figure 1b) indicates the limit after which the regime changes from elastic to plastic one. The structure within this regime no longer present ripples and becomes planar. At this region, we can estimate PG Young's modulus along the *x* direction.

The region (iii) indicates the PG plastic regime of deformations along the *x* direction. PG structures fracture at strains of 14% (10%) at 300 K (900 K). The literature value of the Young's modulus of PG membranes is 800 ± 14 GPa [2]. We found at 300K (900 K), 737 (713) and 808 (847) GPa along *x* and *y* directions, respectively. Our average value for the Young's modulus of PG is 776 GPa, so in good agreement with the values obtained in the literature using different potentials.

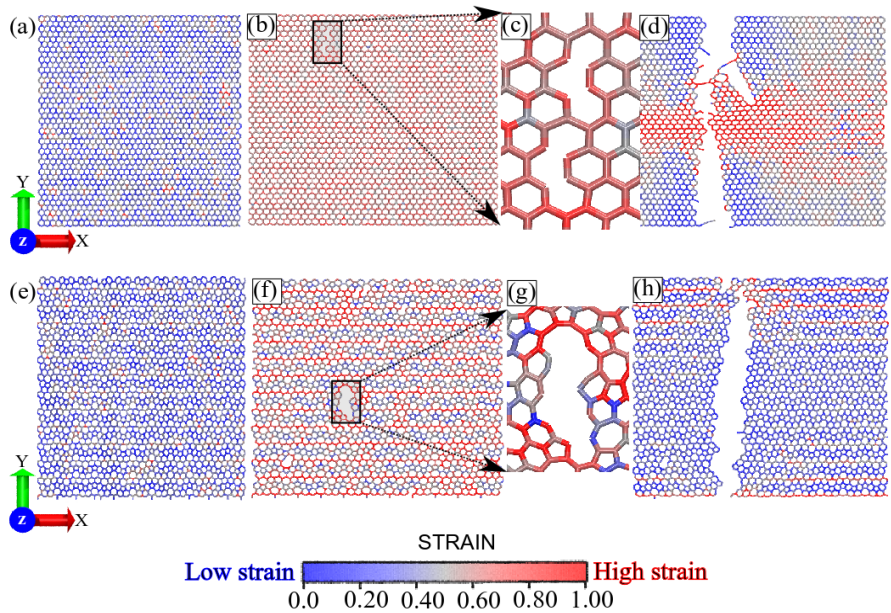

**Figure 2.** MD snapshots for graphene and PG structures, with axial stress applied along the *x* direction at 300 K. (a) Graphene at 0% strain, and; (b) after 9.42% strain, where the beginning of the fracture can be visualized. (c) Zoomed view of the region where fracture starts. (d) Graphene completely fractured at 10.47% strain. (e) PG at 0% strain, and; (f) after 14.10% strain, where the beginning of the fracture can be visualized. (g) ) Zoomed view of the region where fracture starts. (h) PG completely fractured at 14.16% strain. The horizontal bar represents von Mises stress values.

In Figure 2, we present representative MD snapshots of the stretching and fracture dynamics along the *x* direction for graphene PG structures at 300 K. In Figures 2a and 2b, we have a graphene membrane relaxed at 300 K with 0.0% and 9.42% strain, respectively. In Figure 2b, at 9.42% strain, we can see the starting of the fracture process (highlighted in Figure 2c). We can observe here that the first broken carbon-carbon bonds are those that are aligned with the direction of the axial tension when applied along the *x* direction. In Figure 2d, we present a MD snapshot of a completely fractured graphene structure with an approximately linear crack along the direction perpendicular to the applied axial tension. In Figure 2e, we present a PG structure at 0% strain equilibrated at 300K. In Figure 2f, we present a PG structure at 14.10% strain, where we can observe in the small highlighted box, the starting of the fracturing process. In Figure 2h we can observe a completely fractured PG structure. The high concentration of von Mises stress on the C-C bonds that are parallel to the direction of the applied tension results in the observed fracture patterns. As we can see from Figure 2, graphene and PG exhibit similar fracture patterns, but in contrast to graphene in the PG case just one crack propagation was observed.

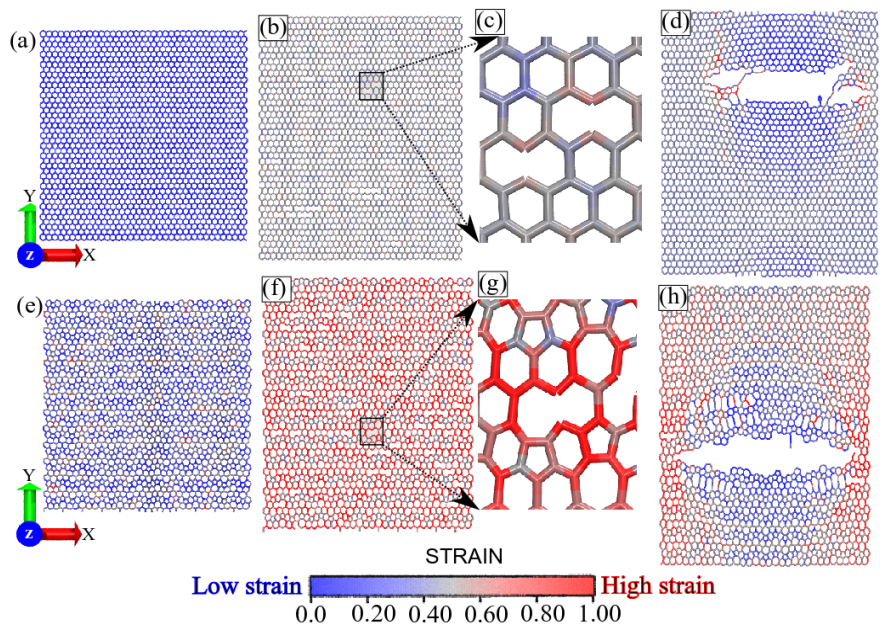

**Figure 3.** Snapshots of the MD simulation for graphene and PG structures with axial stresses applied along the *y* direction at 300 K. (a) Graphene at 0% strain and (b) after 12.9% of strain where the beginning of the fracture is visualized. (c) An enlargement of the fracture beginning. (d) Graphene completely fractured at 12.95% of strain. (e) PG at 0% of strain and (f) after 17.41% of strain, where the beginning of the fracture is visualized. (g) An enlargement of the fracture beginning. (h) PG completely fractured at 17.45% of strain. The horizontal bar represents von Mises stress concentration rates.

In Figure 3, we present the corresponding results for the cases of the stretching applied along the *y* direction Again, we can see that the broken bonds are those that were aligned to the direction of the applied axial tension and PG presents fewer crack propagations in comparison to graphene cases.

## CONCLUSIONS

We have investigated through fully atomistic reactive molecular dynamics simulations the structural and mechanical properties (elastic behavior and fracture dynamics) of graphene and phagraphene (PG) membranes, with axial strain applied separately along the $x$ and $y$ directions. The PG Young's modulus values are about 37% lower than that of graphene, the PG array of 5-6-7 rings results in less dense and softer structure. Interestingly, an unusual anisotropic *pre-elastic* behaviour along the $x$ direction of applied tension was observed at 300 K (not present in the $y$ case nor for 900 K). This is a consequence of the blucked/rippled PG eqilibrium configuration, where it is possible to stretch the structure at low stress. The PG fracture dynamics is similar (fracture starts from bonds alligned parallel to the applied axial strain) to graphene but with fewer crack propations for both $x$ and $y$ cases.

## ACKNOWLEDGMENTS

This work was supported in part by the Brazilian Agencies CAPES, CNPq and FAPESP. The authors thank the Center for Computational Engineering and Sciences at Unicamp for financial support through the FAPESP/CEPID Grant #2013/08293-7. AFF is a fellow of the Brazilian Agency CNPq (#302750/2015-0) and acknowledges support from FAPESP grant #2016/00023-9 and FAEPEX/UNICAMP. ECG acknowledges support from Conselho Nacional de Desenvolvimento Científico e Tecnológico (CNPq) (Process Number 473714/2013-2). AGSF, ECG and JMS acknowledge support from Coordenação de Aperfeiçoamento de Pessoal de Nível Superior (CAPES) through the Science Without Borders program (Project Number A085/2013).